# Phase Separation and Superconductivity in $Fe_{1+x}Te_{0.5}Se_{0.5}$

Vikas Bhatia[a], Efrain E. Rodriguez[b], Nicholas P. Butch[c], Johnpierre Paglione[c] and Mark A. Green[a,b,*]

$Fe_{1+x}Te_{0.5}Se_{0.5}$ is the archetypical iron-based superconductor. Here we show that the superconducting state is controlled by the stacking of its anti-PbO layers, such that homogeneous ordering hinders superconductivity and the highest volume fractions are observed in phase separated structures as evidenced by either a distribution of lattice parameters or microstrain.

Since the discovery of superconductivity in $LaO_{1-x}F_xFeAs$ at 26 K[1] a number of families of iron based high temperature superconductors have been established.[2,3] These include the $ThCr_2Si_2$-type system, $AFe_2As_2$ (A=Ba, Sr, Ca)[4,5] and anti-PbFCl structure of AFeAs (A= Na or Li).[6,7] The same fundamental structural building block in these systems, the tetrahedrally coordinated iron layers, crystallizes without the need of charge balancing cations in FeCh (Ch = Te, Se, S) with the tetragonal anti-PbO structure shown in the inset of Figure 1. In this family, $Fe_{1+x}Te$ at low temperature is monoclinic with a commensurate antiferromagnetic structure for $x \leq 0.12$, which transforms to a structure with orthorhombic symmetry and an incommensurate antiferromagnetic structure at higher amount of interstitial iron, via a phase separated region.[8,9] The critical interstitial iron can be reduced by a post-synthesis reaction with iodine,[10] which can be used to tune the superconductivity.[11] Suppression of the magnetic order and superconductivity can be achieved with either Se ($T_c$ = 14 K)[12] or S($T_c$ = 9 K)[13,14] substitution, but which is associated with considerable inhomogeneity and phase separation.[14] The superconducting transition temperature can be increased considerably to above 30 K on potassium doping[15], leaving a iron vacancy ordered cell[16].

Compositions of $Fe_{1+x}Te_{1-y}Se_y$ close to y ≈ 0.5 have been most widely studied because the chemical pressure on substituting Te with the smaller Se ion, reduces the amount of interstitial iron and greater superconducting volume fractions are observed.[14,17] Here we present, samples of nominal composition, $Fe_{1+x}Te_{0.5}Se_{0.5}$, which have undergone different cooling profiles from the synthesis temperature of 850 ºC. We show that the disorder between the Fe(Te,Se) planes controls the superconductivity in a somewhat counterintuitive manner. The compounds with highest structural homogeneity show little or no superconductivity, whereas those that possess phase separation and variation along the stacking layers show high superconducting volume fractions. These results have widespread implications on the mechanism governing superconductivity.

Samples with nominal composition, $Fe_{1.0}Te_{0.5}Se_{0.5}$, were synthesized by heating stoichiometric quantities of the constituent elements to 425 ºC for 12 hours to begin the reaction, followed by 12 hours at 850 ºC. Various cooling procedures were then adopted that gave different products as evidenced by powder neutron diffraction and SQUID magnetometry, which can be categorized as follows: (1) quenched from above 750 ºC, (2) slow cooled or at the natural cooling rate of the furnace from the synthesis temperature and then quenched from a lower temperature, (3) cooled at the natural temperature rate of the furnace or relatively quickly (≈ 2º/min), (4) cooled over an extended period of up to 2 weeks. Powder neutron diffraction was obtained at T = 4 K using the BT1 diffractometer at NIST (Ge311, λ = 2.0782 Å). These measurements showed that the samples from the different cooling procedures possessed highly varied interlayer stacking and anion composition distribution. Figure 1 shows the full width at half maximum (FWHM) as a function of two-theta of a sample cooled over 14 days. Reflections with no $c$ contribution, such as [110], [200] and [220], are close to the resolution function of the instrument, whereas those with contribution along c are broadened, for example the [001] is ≈ 60 % and the [003] is ≈ 100 % wider than the resolution function.

Two tactics were employed to describe this microstructure within the Rietveld refinement. Firstly, three quartic form anisotropic strain-broadening terms, $S_{400}$, $S_{004}$ and $S_{112}$, were employed. Inclusion of additional terms did not improve the fit in any of the neutron diffraction patterns. Secondly, these fits were compared with multi-phase refinements with up to four phases. Figure S1 (supplementary) shows the best fit obtained by Rietveld refinements of a typical powder neutron diffraction pattern for each of the heating profiles.

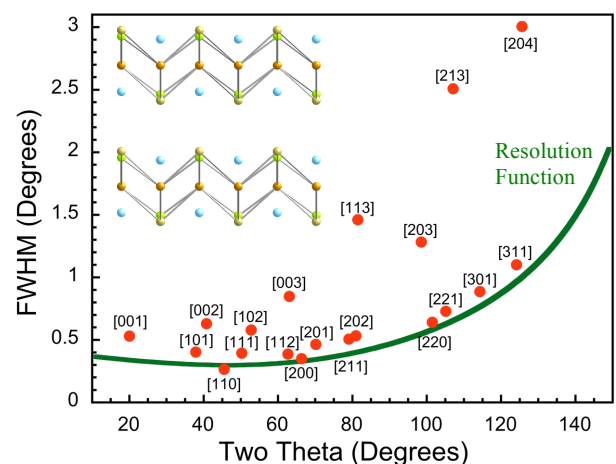

**Figure 1** FWHM of $Fe_{1.0}Te_{0.5}Se_{0.5}$ that has been cooled from 750 ºC to room temperature over 14 days, showing significant broadening to reflections with c contribution. Inset shows the anti-PbO structure of $Fe_{1+x}Te_{1-y}Se_y$ showing the van der Waals separated layers of Fe (orange) bonded to the Te (yellow), Se (green) split site. The interstitial Fe (blue) takes up a square pyramidal site with the (Te, Se) plane.

Cooling profile 1 (high temperature quench) gave phases that have a considerable amount of hexagonal Fe(Te,Se), which has the NiAs structure, rather the tetragonal anti-PbO phase. The sample that neutron diffraction was performed on was best fitted with hexagonal Fe(Te,Se) and two phases of the anti-PbO Fe(Te,Se); one Te-rich phase with $a = b = 3.79793(6)$ Å and $c = 6.0343(2)$ Å, and a Se-rich phase with $a = b = 3.7906(2)$ Å and $c = 5.8748(5)$ Å. The ratio of hexagonal : Te-rich : Se-rich phases was 22(1) : 48(1) : 30(1), respectively. Both of the anti-PbO phases, refined with isotropic Lorentzian broadening that was above the resolution function, but no significant anisotropic broadening was observed. Figure 2 shows the two distinct (001) reflections at ≈ 20° two-theta and the (112)/(003) doublet at ≈ 63° two-theta that arises from the well defined Te-rich and Se-rich anti-PbO structures.

Cooling profile 2, allowed the sample initially to slow cool, but then quenched from a much reduced temperature than cooling profile 1, at around 440 °C (low temperature quench), as demonstrated to be successful in the synthesis of FeSe[18]. This procedure significantly reduces the amount of hexagonal phase that is the thermodynamic stable phase at high temperature. In this procedure, two anti-PbO phases were present, one Te-rich the other Se- rich, but these phases are different to those seen in the high temperature quench. The Te-rich phase with $a = b = 3.79859(5)$ Å and $c = 5.9792(1)$ Å, has a peak shape at the resolution function of the neutron powder diffraction experiment, implying no broadening associated with strain or a distribution of lattice parameters. In contrast, the Se-rich phase, with $a = b = 3.7910(1)$ Å and $c = 5.861(1)$ Å, has significant broadening along the $c$ direction. A comparison between the strain broadening and multiphase reflections showed that the anisotropic strain gave a significantly better fit as a result of the homogeneity of the broadening and even distribution in the $c$ parameter, which is shown in Figure 2. The distribution of c parameter is estimated to be from 5.62 to 6.11 Å from the width of the (001) reflection; this corresponds to a very broad distribution over near the entire Fe(Te, Se) solid solution. The ratio of hexagonal : Te-rich : Se-rich phases was now refined to be 5(1) : 51(2) : 44(2) : 5(1), respectively.

The third cooling profile (12 hour cool) was based on more typical cooling rates that brought the furnace to ambient temperature in 7 – 12 hours. As can be seen in Figure 2, the resulting set of (001) and (112)/(003) reflections are very different to the quenching methods; now the peak shape profile is very close the resolution function of the instrument implying a much narrower distribution of lattice parameters and well defined stoichiometry that refined to be FeTe$_{0.57(2)}$Se$_{0.43(2)}$, with $a = b = 3.79510(3)$ Å and $c = 5.9194(1)$ Å. Note, the interstitial iron occupancy refined to zero in this and all other refinements. Application of anisotropic strain broadening did lower the resulting R-factor from 8.52 to 6.42 %, but the extent of the broadening is significantly lower than previous samples. For example, the $S_{004}$ broadening term was 2.3(1) in the 12 hour cool sample, compared with 125 (5) for the low temperature quench.

In contrast, cooling profile 4 (2 week cool) gave samples similar to cooling profile 2, with substantial broadening of the reflections along the $c$ direction. However in this profile, the broadened reflections take on distinct features suggesting the presence of diffracting domains rather than the broad distribution of lattice parameters found for the low temperature quench procedure. Rietveld refinement using the neutron diffraction data was much improved employing a multi-phase refinement over a strain model fit. The R-factor fell from 17.1% for a single phase model without strain broadening to 8.98% for a model with $S_{004}$ broadening of 22.9(8); significantly less broadening compared to the disorder imparted on the sample as a result of a quench from low temperature. A two-phase model gave an R-factor of 10.7 %, which is similar to the single-phase model with strain. However this fell to 5.4 % when three phase model were employed. Figure 2 shows how these three phases accurately model the distinct broadening to the (001) reflection as well as the (112)/(003) doublet. The lattice parameters for the three phases were distributed from 3.7805 (5) – 3.79391(9) Å for a = b, and 5.727(2) – 5.9616(3) Å for the $c$ parameter, which is a very significant distribution of lattice parameters but which is still smaller than the values resulting in a quench from low temperature.

To evaluate the consequences of this microstrain and phase separation on the superconducting properties, a series of magnetization measurements were performed on a commercial SQUID magnetometer, which are summarized in Figure 3. The samples that showed no (00l) broadening and possess a homogeneous distribution of *c* parameters as a result of the 12 hour cooling procedure, are surprisingly not bulk superconductors, and contain less than 10 % superconducting volume fraction. An increase to around 40 % is seen in the high temperature quenched samples that contains two tetragonal Fe(Te, Se) phases and isotropic broadening of the peak shape. Bulk superconductivity at or above 80% superconducting volume fraction is observed in both the low temperature quenched and the extended cooling procedure (2 week cool). Both of these phases have considerable amount of lattice parameter distribution and strain from the establishment of Te and Se rich regions.

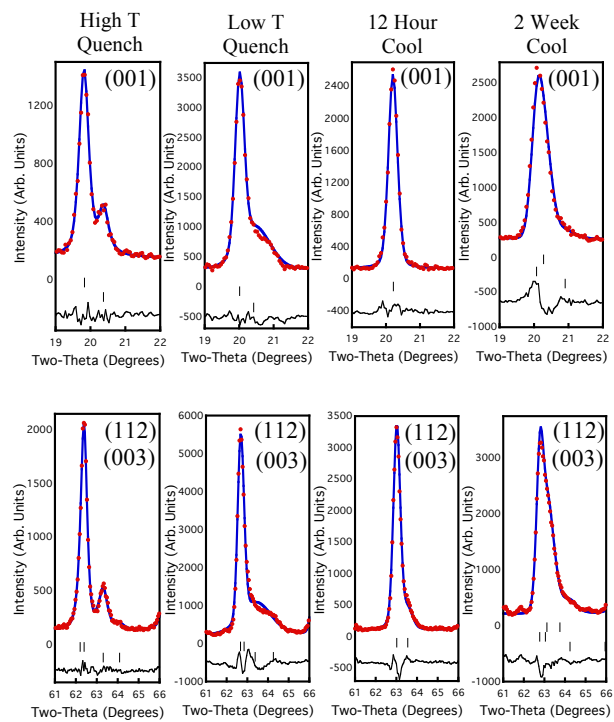

**Figure 2** The (001) reflection (top) and the (112)/(003) doublet (bottom) for four distinct thermal treatments showing contrast between the single phase product obtained from a 12 hour cooling procedure, compared with combinations of anisotropic broadening and phase separation present when using other cooling procedures.

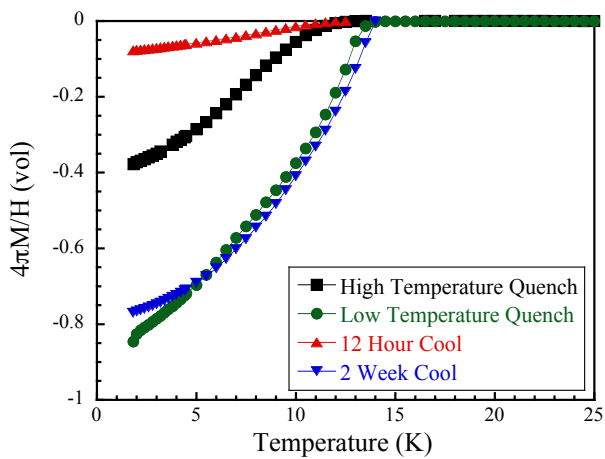

**Figure 3** Superconducting volume fractions for the four samples synthesized from distinct cooling profiles, demonstrating very contrasting superconducting properties.

## Conclusions

In a systematic investigation into inhomogeneity and superconductivity in the anti-PbO structure of nominal composition, $Fe_{1+x}Te_{0.5}Se_{0.5}$, a surprising inverse relationship is established. The samples with considerable inhomogeneity through the presence of phase separation and lattice strain, actually possess higher superconducting volume fractions than those whose neutron diffraction peak shapes confirm homogeneity above the measurement capability of the high resolution powder neutron diffractometer. This is particularly surprising as the composition of the 12 hour cool sample used for neutron diffraction, $FeTe_{0.57(2)}Se_{0.43(2)}$, is well-established in the literature as optimal for superconductivity. Although recent work on $Fe_{1+x}Te_{0.7}Se_{0.3}$ confirmed that superconductivity is robust across a very wide range of compositions[11], this work establish that inhomogeneity is the critical parameter. Much of the work on this family of superconductors has been performed on single crystals, which often has considerably more disorder than powdered samples as a result of the nature of the growth mechanism. The 2 week cooling procedure would be comparable to the timescale for growing single crystals through the Bridgeman technique.

Within the Fe(Te, Se, S) series, and other systems where superconductivity is induced by isovalent substitution, such as $Ba(Fe, Co)_2As_2$, it has not been clear why the electronic properties are so very different across the solid solution. This is in stark contrast to chemistry of the cuprates that rely on optimizing the copper oxidation state. However, in all of these systems with isovalent substitution, the different cation radii incorporate a considerable amount of addition inhomogeneity that is here demonstrated to be a critical factor. In addition, the fact that high superconducting volume fractions arise from both compounds that possess a very wide range of compositions (low temperature quench) and compounds where growth of large domains of better-defined Te-rich and Se- rich domains is performed, suggests that the presence of the phase separation is more important than the details of its structure.

## Notes and references

[a] *Department of Materials Science and Engineering, University of Maryland, College Park, MD, 20742 Email: mark.green@nist.gov*
[b] *NIST Center for Neutron Research, National Institute of Standards and technology, 100 Bureau Drive, Gaithersburg, MD 20899*
[c] *Center for Nanophysics and Advance Materials, University of Maryland, College Park, MD 20742*

† Electronic Supplementary Information (ESI) available: See DOI: 10.1039/b000000x/